\documentclass[12pt]{article}
\usepackage{graphicx}
\usepackage{url}

\usepackage{times}
\usepackage{times}
\newenvironment{sciabstract}{%
\begin{quote} \bf}
{\end{quote}}

\newcounter{lastnote}

\def\figwidth{3.4in}

\title{Quality versus quantity in scientific impact}

\author{Jasleen Kaur,$^{1}$ Emilio Ferrara,$^{1}$ Filippo Menczer,$^1$\\ Alessandro Flammini,$^1$ Filippo Radicchi,$^1$
\\
\normalsize{$^1$ Center for Complex Networks and Systems Research}\\
\normalsize{School of Informatics and Computing}\\
\normalsize{Indiana University, Bloomington, USA}
}

\date{}
\begin{document}
\baselineskip24pt
\maketitle



\maketitle

\begin{sciabstract} 
Citation metrics are becoming pervasive in the quantitative evaluation 
of scholars, journals and institutions. 
More then ever before, hiring, promotion, and funding decisions 
rely on a variety of impact metrics that cannot disentangle quality 
from quantity of scientific output, and are  
biased by factors such as discipline and academic age. 
Biases affecting the evaluation of single papers 
are compounded when one aggregates citation-based metrics across
an entire publication record. It is not trivial to compare the quality 
of two scholars that during their careers have published at different 
rates in different disciplines in different periods of time. 
We propose a novel solution based on the generation of a statistical 
baseline specifically tailored on the academic profile of each researcher. 
Our method can decouple the roles of quantity and quality of publications 
to explain how a certain level of impact is achieved. The method is flexible enough 
to allow for the evaluation of, and fair comparison among, arbitrary collections 
of papers --- scholar publication records, journals, and entire institutions; 
and can be extended to simultaneously suppresses any source of bias. 
We show that our method can capture the quality of the work of Nobel laureates 
irrespective of number of publications, academic age, and discipline, 
even when traditional metrics indicate low impact in absolute terms. 
We further apply our methodology to almost a million scholars and over 
six thousand journals to measure the impact that cannot be explained 
by the volume of publications alone. 
\end{sciabstract}



\section*{Introduction}

The interest in measuring scientific impact is no longer restricted to bibliometrics specialists, but extends to the entire scientific community. 
Many aspects of academic life are influenced by impact metrics: from the desire to publish in high-impact journals~\cite{calcagno2012flows}, to hiring, promotion and funding decisions~\cite{bornmann2006selecting}, and department or university rankings~\cite{davis1984faculty, liu2005academic}. 
Although the idea of measuring scientific impact is laudable, several fundamental aspects in the current evaluation methods are problematic; the use of existing citation-based metrics as proxies for ``true'' scientific quality of publications or scholars in practical contexts is often unsatisfactory~\cite{adler2009citation}, or worse, misleading~\cite{Alberts17052013,Nature-JIF}. 
Comparisons among scholars, journals, and organizations are meaningful only if one takes into account the proper contextual information, such as discipline, academic age, publication and citation patterns.

Some of these issues can be addressed at the level of individual publications. 
Two important factors affecting the citations of an article are discipline and age. Once papers are divided into homogeneous sets according to these features, the populations within these classes can be used as baselines. One intuitive approach is that of assigning papers to citation percentiles~\cite{leydesdorff2011turning}.
Another possibility is to leverage the universality of citation distributions to measure relative citation counts~\cite{radicchi2008universality, radicchi2012reverse}.
The situation, however, becomes more challenging when we try to assess the quality of \emph{aggregate} entities such as scholars, journals, or organizations.
There have been several attempts to measure the impact of such entities that rely on aggregating across all the papers that can be attributed to the entity. Of course, the biases that affect the evaluation of individual papers are amplified when these aggregate measures are considered. Most impact metrics have been shown to be strongly biased by multiple factors when authors are considered~\cite{alonso2009h, duch2012possible, radicchi2013analysis, KRM} and corrections to mitigate biases
due to discipline, multiple authors, and academic age have been proposed~\cite{batista2006possible, sidiropoulos2007generalized, Schreiber2008, waltman2011towards}. Unfortunately none of these corrections is effective against the whole spectrum of potential biases~\cite{KRM}. 

The biases of impact metrics for researchers cannot be addressed with the same classification-based approach as for individual publications; scholars cannot be simply divided into categories that are simultaneously homogeneous for academic age and scientific discipline. 
First, it is not clear whether age should be quantified in terms of academic years of activity or total number of publications.
Fixing only one of these two constraints would lead to a large variability for the other quantity.
Accounting for both, instead, would produce sparsely populated categories of no practical use.
Second, many researchers work on a range of different topics and in multiple disciplines~\cite{Sun-SDS}, or change their research interests during their careers. Therefore, reducing a scholar's research to a restrictive scientific subject container makes little sense. 
Also here, focusing only on scholars who are involved in exactly the same set of topics would generate very sparse categories. 
The situation only worsens if one simultaneously takes into account 
age, disciplines, and their intricate longitudinal combinations. 
We propose a novel statistical method that addresses these issues by evaluating  quality in the proper context.

\section*{Materials and Methods}

\begin{figure}
\centerline{\includegraphics[width=\figwidth]{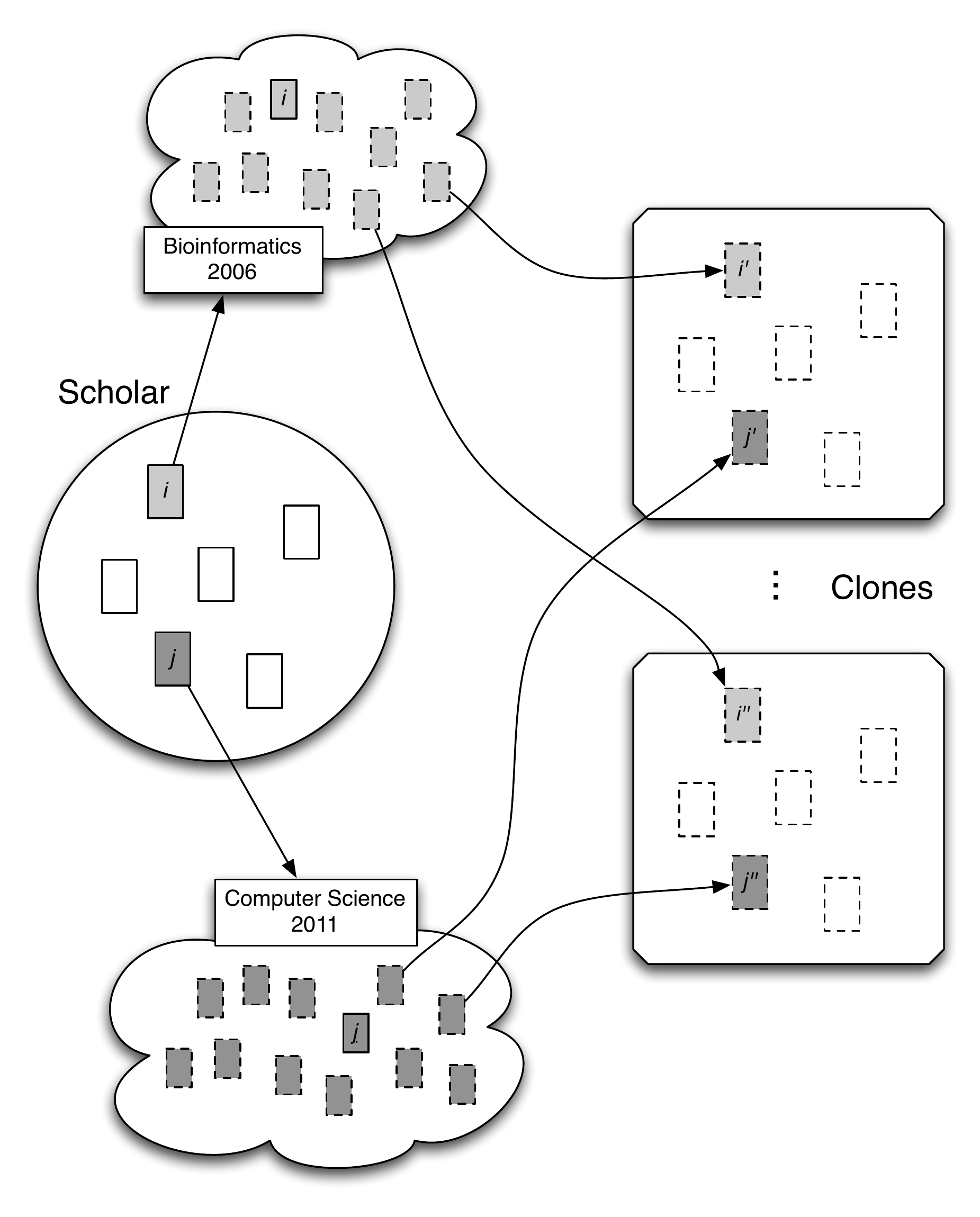}}
\caption{Schematic illustration of the resampling technique used 
to generate the publication records of a scholar's clones. A
scholar's paper $i$ is replaced in a clone by a randomly 
selected paper $i'$, published in the same year and 
in the same subject category. Similarly the paper
is replaced by another paper $i''$, from the same set, in a different clone.
The same resampling is applied to each paper for each clone.}
\label{fig1}
\end{figure}

Our approach starts from an aggregate set of papers. While this can apply to scholars, journals, or institutions, let us illustrate it in the case of a researcher. 
The idea is to generate a statistical baseline specifically tailored on the academic profile of the scholar; the term of comparison is not given by other individuals, but rather by artificial copies of that scholar. 
Each copy, or \emph{clone}, has a publication record with identical publication years and subject categories as the researcher under observation. However, the citation profile is resampled: the number of citations of each paper is replaced by that of a paper randomly selected among those published in the same year and in the same discipline. The cloning procedure is illustrated in Figure~\ref{fig1}.  

In essence, a clone encodes an academic trajectory that in number of papers, their publication years, and topics exactly corresponds to that of the scholar being cloned. 
One can compute any citation-based impact metric for a clone, given its citation profile. From a population of clones associated with a researcher profile, one can estimate the likelihood that the scholar's measured impact could be observed by chance, given her publication history. 
Since the publication history includes the number of publications, this procedure deals with the biases that affect this number, such as academic age and disciplinary publication practices. In other words, the procedure decouples quantity and quality, allowing to ask whether a certain level of impact can be explained by quantity alone, or an additional ingredient of \emph{scientific quality} is necessary. 

More specifically, consider a researcher $r$ who published $N_r$ papers, in specific years \linebreak $\{y_1, y_2, \ldots, y_{N_r}\}$ and disciplines $\{s_1, s_2, \ldots, s_{N_r}\}$, that have received certain numbers of citations $\{c_1, c_2, \ldots, c_{N_r}\}$, where $y_i$, $s_i$ and $c_i$ indicate respectively the year of publication, the subject category, and the total number of citations accumulated by the $i$-th paper. Any citation-based impact metric for  $r$ can be calculated using this information, including simple ones, like total or average number of citations, or more sophisticated ones like the $h$-index~\cite{hirsch2005index}. Let $m_r$ be the observed score of the metric $m$ for researcher $r$. A clone of $r$ is generated by preserving the years and subject categories of the entire publication record of $r$, but replacing the number of citations $c_i$ accumulated by every paper $i$ with that of another paper randomly selected from the set of articles published in year $y_i$ in subject category $s_i$. 
Once a clone is generated, we measure the value $m'_r$ of the same impact metric $m$ on its profile. After repeating this operation $T$ times on as many independently generated clones, we compute the \emph{quality score} $q$ as the fraction of times that $m_r \geq m'_r$. We also compute the \emph{standard score} $z_r = (m_r - \overline{m}_r) / \sigma_r$, where $\overline{m}_r$ and $\sigma_r$ are the mean the standard deviation of $m$ over the population of $r$'s clones. Our numerical results are obtained using $T=1000$.

\subsection*{Disciplines and publication venues}

The cloning method relies on the classification of articles in subject categories.  The discipline label $s_i$ for a paper $i$ may not be directly available in the data, but can be inferred by its publication venue $v_i$. Mapping venues to disciplines and vice versa requires a Bayesian framework to properly account for the many-to-many relationship between venues and subject categories.  
In practice, given a paper published in venue $v$ in year $y$, we wish to replace its number of citations with those of another paper chosen at random among all publications in venue $v'$ and year $y$. To select $v'$ we need to estimate the conditional probability $P_y(v'|v)$ that a paper published in year $y$ and in venue $v$ could have been potentially published in venue $v'$. 
Let us encode the classification of venues in subject categories via a matrix $B$, so that element $B_{vs}=1$ if venue $v$ is classified in subject category $s$, and $B_{vs}=0$, otherwise. The probability that a randomly selected paper published in year $y$ belongs to venue $v$ is defined as
\begin{equation}
P_y(v) = \frac{N_y(v)}{\sum_{u} N_y(u)}
\label{eq:pj}
\end{equation}
where $N_y(v)$ represents the total number of papers published in venue $v$ in year $y$. The probability that, given the venue $v$
of publication, a paper belongs to category $s$ is given by
\begin{equation}
P_y(s|v) =
 \frac{N_y(s) B_{vs}}{\sum_{s'} N_y(s') B_{vs'}} 
 \label{eq:sj}
\end{equation}
where $N_y(s) = \sum_u B_{us} N_y(u)$ represents the total number of papers published in year $y$ in venues belonging to category $s$. We can now formulate the probability that a randomly selected paper published in year $y$ belongs to category $s$:
\begin{equation}
P_y(s) = \sum_{v} P_y(v) P_y(s|v) \;.
\label{eq:ps}
\end{equation}
We then use Bayes' theorem to calculate the probability that a paper in category $s$ is
published in venue $v$:
\begin{equation}
P_y(v|s) = \frac{P_y(s|v) P_y(v)}{P_y(s)} \;.
\label{eq:js}
\end{equation}
Finally, we have all the elements needed to compute
\begin{equation}
P_y(v'|v) = \sum_s P_y(v'|s) \, P_y(s|v) \;. 
\label{eq:jj}
\end{equation}

\subsection*{Bibliographic data}

To implement our resampling procedure
we consider all papers indexed in the Web of Science database published between  1970 and 2011~\cite{wos}. These amount to 22,088,262 records classified as ``articles,'' ``proceedings papers,'' ``notes,'' or ``letters,'' and cover publications in 9,696 science and social sciences venues. 
Based on the publication venue, we associate each article to one or more of 226 subject categories, as defined in the Journal Citation Reports database~\cite{jcr}. The number of citations accumulated by each publication in our dataset 
was retrieved in March--April 2012.
Authors are identified on the basis of first and middle name initials and full last name. Although we did not implement any disambiguation algorithm, we expect errors in author identification to account for less than 5\% of the records~\cite{radicchi2009diffusion}. We further restrict our attention only to authors with a publication record of at least 10 and at most 500 articles, filtering out many ambiguous names. The subsequent analysis is based on 996,288 author records matching these criteria.

\section*{Results}

\subsection*{General properties of the quality score}

Whereas the procedure described in the
methods section can be applied to any citation-based 
metric, let us first consider the $h$-index~\cite{hirsch2005index}, 
widely used to represent the productivity and impact of 
researchers with a single number. 
In our analysis, we estimate the probability $q$ 
that a clone's $h$ is less than that of the corresponding scholar. 
Values of  $q$ close  to $0$ mean that the scholar's impact (as measured by $h$) is much smaller than one would expect from her publication profile (number of papers, and relative publication years and disciplines).  
Conversely, $q$ close to $1$ suggests that the author produced publications of consistently high quality. 
  
\begin{figure}
\centerline{\includegraphics[width=\figwidth]{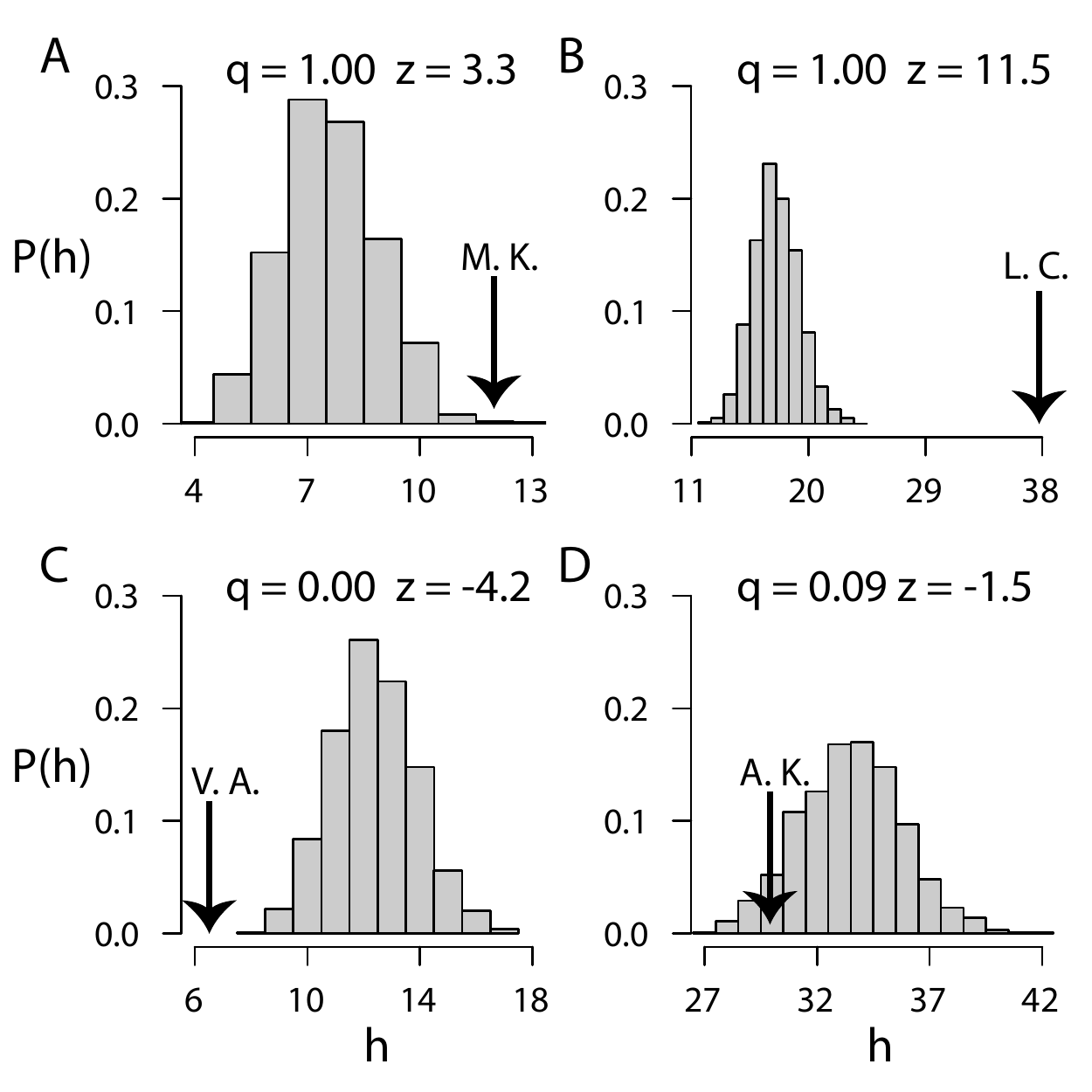}}
\caption{Distributions of $h$ for clones of four mathematicians, yielding different values of $q$ and $z$. Arrows indicate actual $h$ values. An author can have low $h$ and high $q$ (A), high $h$ and high $q$ (B), low $h$ and low $q$ (C), or high $h$ and low $q$ (D). The two scholars with high $q$ are a Fields medalist and a Wolf Prize recipient.}
\label{fig2}
\end{figure}

Figure~\ref{fig2} shows the relationship between the $h$ 
value of four authors and those of their clones, 
yielding different values of $q$. 
In general, there is not a strict correspondence between the values of $h$ and $q$.
A high value of $q$ is indicative of high quality even when the scholar's $h$ is not high in absolute terms, as illustrated by the Fields medalist 
represented in Figure~\ref{fig2}A.
Conversely, a high $h$ does not necessarily imply high quality; most of the clones 
in Figure~\ref{fig2}D have higher $h$ than the scholar. 
The distribution of $h$ values for the clones is
in general compatible with a bell-shaped distribution. This observation supports
our use of the standard score $z$ as a related measure of scientific 
quality. The scores $q$ and $z$ convey similar
information, but $z$ provides higher
resolution than $q$, especially for extremely 
low or high values of $q$. For example, the quality scores of the 
scholars of Figures~\ref{fig2}A and \ref{fig2}B are indistinguishable ($q=1$)
on the basis of the $T=1000$ clones produced. Their standard scores, however, 
provide a basis for finer discrimination ($z=3.3$ and $z=11.5$ respectively).

\begin{figure}
\centerline{\includegraphics[width=\figwidth]{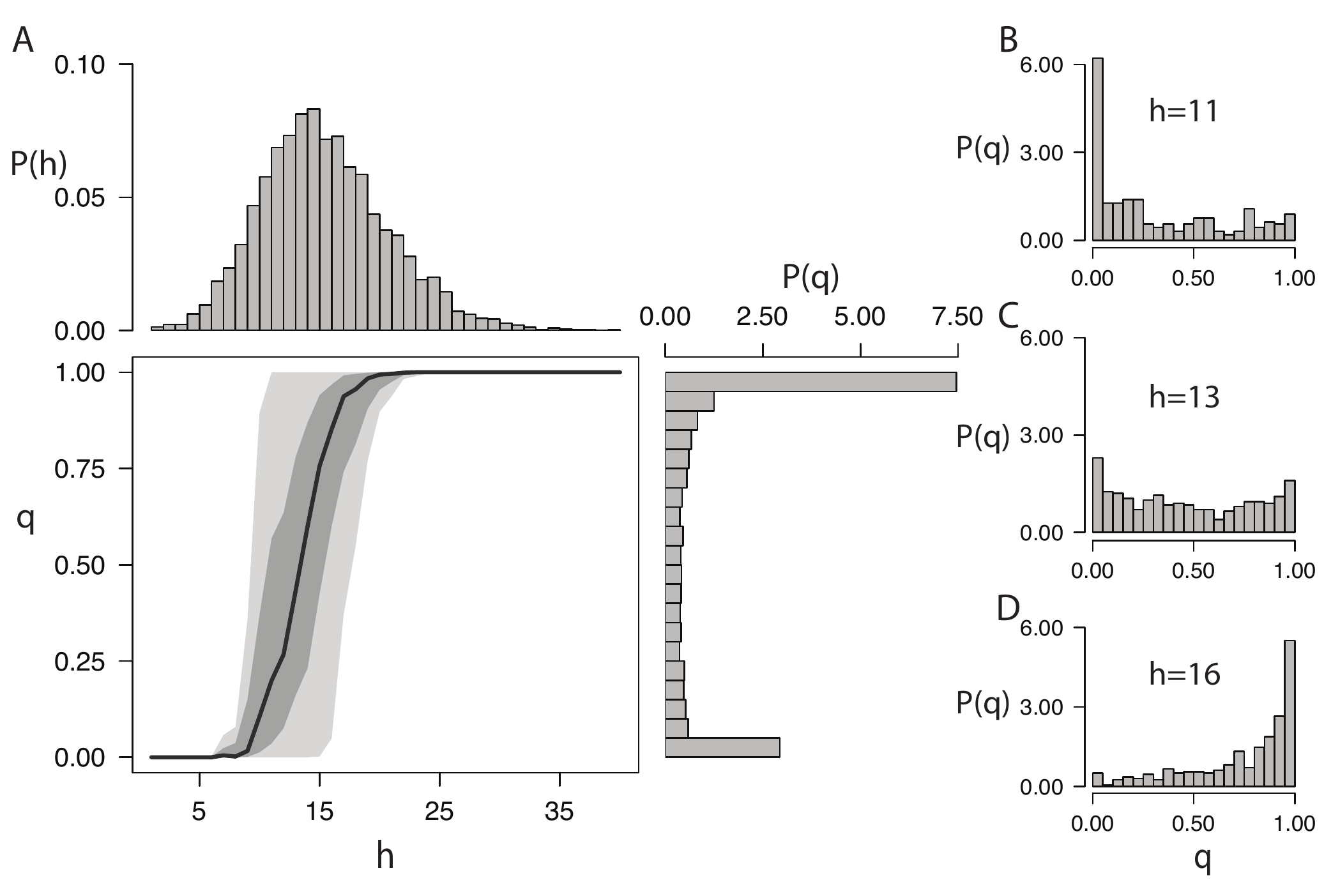}}
\caption{Relationship between $h$ and $q$ for 4,912 authors of 50 papers each. (A) Sharp transition of $q$ around the critical value $h_c \simeq 13$. The black line represents the median $q$, and the gray areas represent the 50\% and 95\% confidence intervals. The distributions of $h$ (top) and $q$ (right) are also shown. Plots (B-D) show the distributions of $q$ values for $h=11,13,16$.}
\label{fig3}
\end{figure}

General properties of the relation between $h$ and $q$ emerge when we consider the entire dataset. Figure~\ref{fig3} shows that, as expected, $q\rightarrow0$ and $q\rightarrow1$ for small  and large values of $h$, respectively. If we restrict ourselves to considering only authors with a fixed number of papers $N$, the transition between these two extremes is in general sharp and located at a critical value $h_c$ that depends on $N$. The great majority of authors with $h < h_c$ have very low $q$, while most authors with $h> h_c$ have very high $q$. Overall, more than half of the researchers have extreme values of $q$: about $22.5\%$ have $q\approx0$, and $30\%$ have $q\approx1$. 

\subsection*{Critical lines of high quality}

\begin{figure}
\centerline{\includegraphics[width=\figwidth]{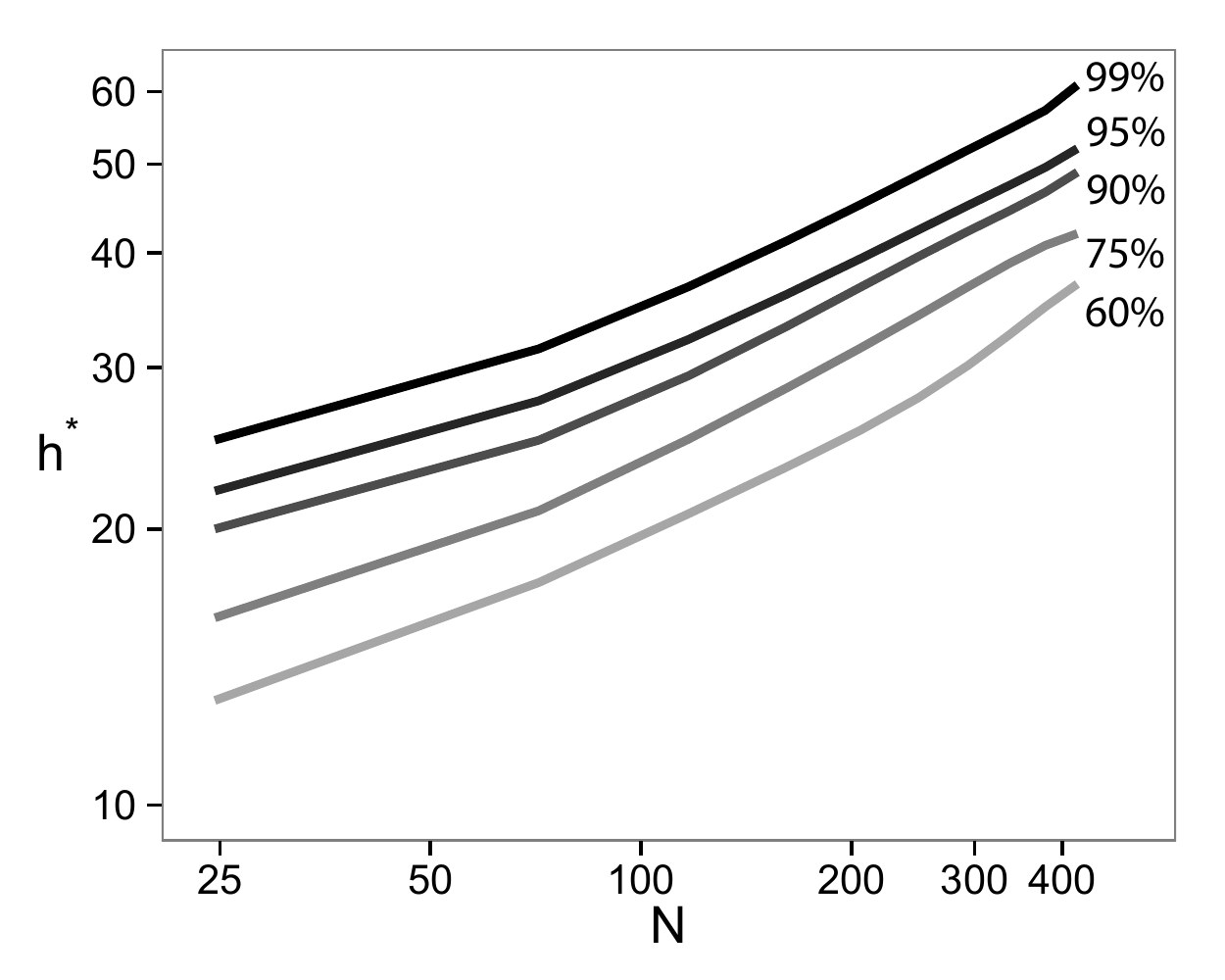}}
\caption{Career phase diagram. Relationship between  $h^*$ and the number of papers $N$ for 996,288 authors. Lines represent the values of $h$  above which a fraction $\delta$ of authors in our dataset have $q$ value larger than 0.95, for $\delta =$ 0.60, 0.75, 0.90, 0.95, and 0.99.} 
\label{fig4}
\end{figure}

What is the value of $h$ necessary to support a case of high scientific quality, given one's productivity (quantity)? The previous analysis suggests that such a value can be determined with some accuracy. Next we describe a procedure that can be employed to answer this question empirically. 
First, we bin authors according to the number of their published papers, $N$.
For each bin, we then determine the value $h^*(N, \delta)$ 
defined by $P(q > 0.95 \;|\; h \geq h^* , N) = \delta$.
$h^*(N, \delta)$ represents the value
of $h$  above which a fraction $\delta$ of scholars 
have a $q$ value larger than $0.95$. 
In Figure~\ref{fig4} we draw $h^*(N, \delta)$ as a function of $N$
for several values of $\delta$. These phase diagrams separated by    
$h^*(N, \delta)$ can be interpreted as critical lines in the career 
trajectory of a scholar. We argue that impact characterized by $h \geq h^*(N, \delta=0.95)$ 
provides strong evidence of high scientific quality.

\subsection*{Validating scientific quality}

\begin{figure}
\centerline{\includegraphics[width=\figwidth]{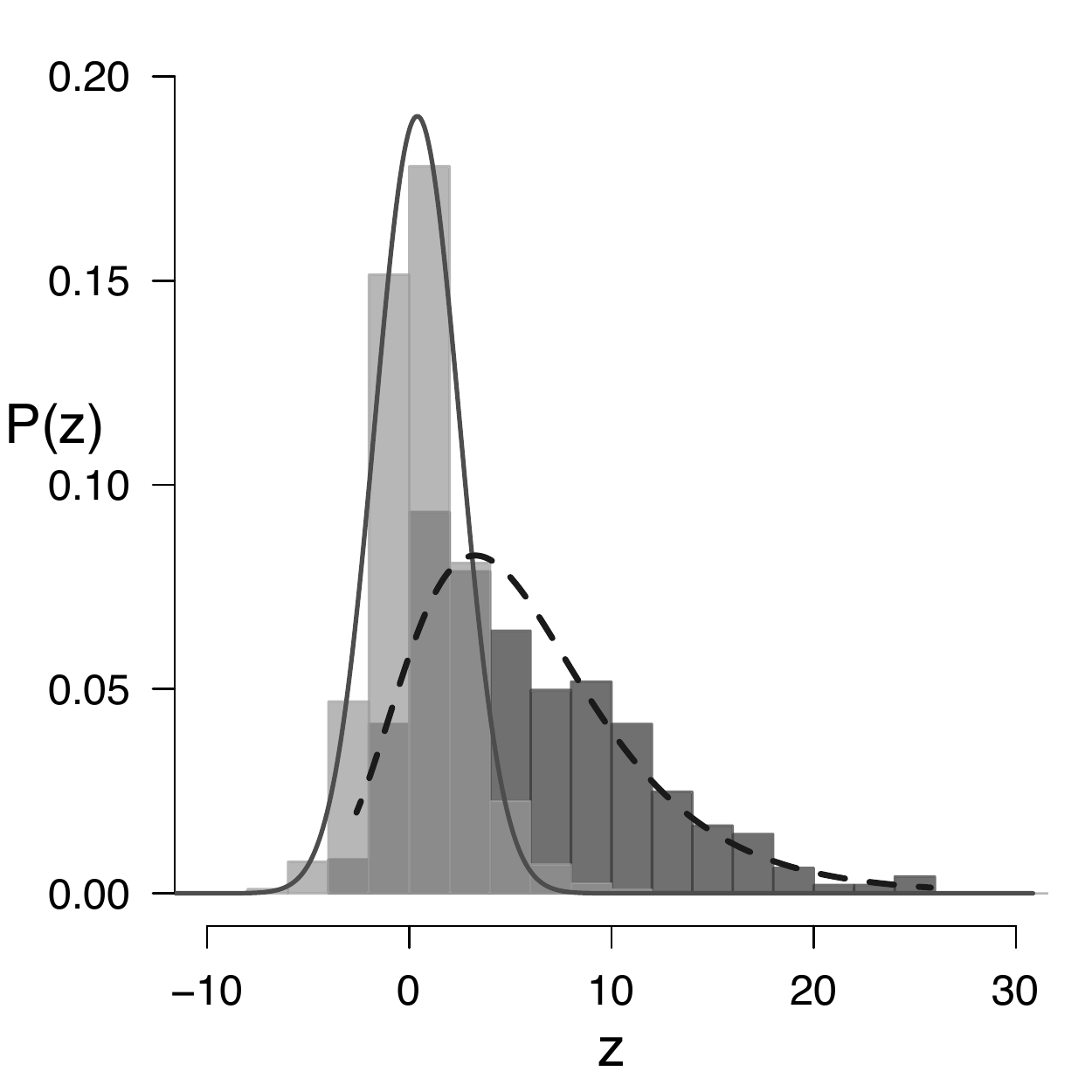}}
\caption{Distribution of $z$ for all scholars in our dataset (light gray). 
$P(z)$ is reasonably well fitted by a normal distribution
with mean $\overline{z} = 0.4 \pm 0.1$ 
and standard deviation $\sigma_z = 2.0 \pm 0.1$ (black line).
The distribution of the entire population is compared
with the one of Nobel laureates (dark gray). 
The latter is compatible with a Gumbel distribution with parameters 
$\mu = 3.5 \pm 0.1$ and $\beta = 4.3 \pm 0.1$ (dashed line).}
\label{fig5}
\end{figure}

As already mentioned, $q$ does not provide
a sufficient resolution at the extremes;
exceptionally good scholars, for example, 
all have $q \approx 1$. It is useful therefore to consider $z$ scores. 
As Figure~\ref{fig5} shows, $P(z)$ can be
fitted  relatively well by a normal distribution. 
Overall, about the $58\%$ of the scholars have $z > 0$.
Researchers with $z<-1$ and $z<-2$ amount 
to $24\%$ and $11\%$ of the population, respectively.
Those with $z>1$ and $z>2$ represent  
$39\%$ and $23\%$ of the sample, respectively.

To test the ability of our quality score  
to recognize ``true'' scientific excellence we
consider all Nobel laureates in the period $1970-2013$.
This set represents a small but ideal benchmark to check 
the validity of our method.  
The selection of Nobel recipients in fact does not depend
on citations: laureates are identified each year
by large committees, often with the help of the entire scientific community. 
The distribution of $z$ values for Nobel laureates
is also shown in Figure~\ref{fig5}. 
Many Nobel recipients have very high $z$ scores, reflecting the exceptional scientific quality revealed by their publication profiles. Their $z$ distribution  can be fitted well by a Gumbel distribution, which, interestingly, is often adopted to describe the statistics of extremes events~\cite{gumbel1958statistics}. 
At the same time, we note that a few Nobel laureates have low $z$ scores.
In some cases this is the result of ambiguous names leading to profiles
with inflated number of publications. For example, S.C.C.~Ting (Physics, 1976) 
appears to have $z=-1.5$. Upon inspection we find that his publication record is composed of 655 publications, the majority of which are attributable to homonymous authors. 
As already noted for the general population of scientists, high values
of $z$ do not necessarily correspond to high $h$.
In Table~\ref{tab1} we list a few examples of Nobel laureates in
various disciplines and years. Among the majority of laureates with 
high $h$ values, we find a few, such as R.H. Coase (Economics, 1991) and 
R. Furchgott (Medicine, 1998), with relatively low $h$ values.  
In these 
cases, the $z$ scores 
is a more reliable indicator of exceptional scientific quality 
compared to $h$. 
We recognize that the metrics $q$ and $z$ of scientific quality, 
when applied to Nobel laureates, could be in principle affected by
the boost in the number of citations typically observed
after the award~\cite{mazloumian2011citation}. 
Nevertheless, even when we focus on 2012 Nobel recipients only, 
we find consistently high $z$ values, on occasion associated 
with relatively low $h$.

\subsection*{Analysis of scientific journals}

\begin{figure}
\centerline{\includegraphics[width=\figwidth]{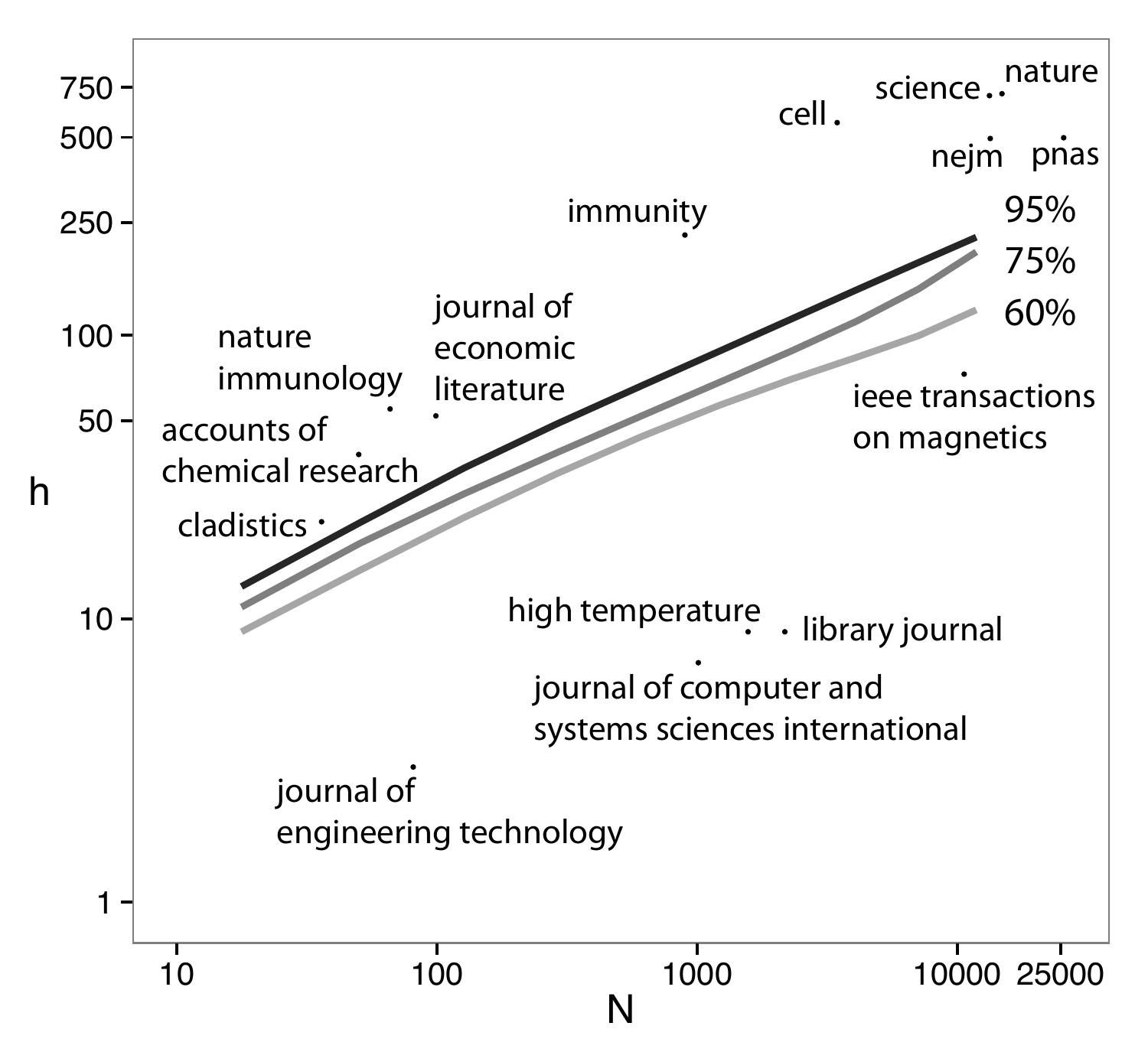}}
\caption{Journal phase diagram. Relationship between $h^*$ and $N$ for 6,129 journals. We consider only papers published in the period 1991--2000. The lines represent different values of $\delta$. Several examples of journals are displayed.}
\label{fig6}
\end{figure}

Although we focused on scholars until now, our general procedure is easily 
applicable to other aggregate entities, such as journals, with essentially no modification. 
Figure~\ref{fig6} summarizes the results obtained for journals. 
We consider all publications in the
period 1991--2000 and use $h$ as impact metric for 
journals~\cite{braun2006}. We then apply our statistical procedure
to calculate the critical $h^*(N, \delta)$ as described above.
Specific examples of academic journals are marked in the 
diagram.  Journals with large numbers of papers and high impact, such as
\textit{Nature}, \textit{Science}, \textit{PNAS},
\textit{Cell}, and the \textit{New England Journal of Medicine}, 
are well above the critical line $h^*(N,\delta=0.95)$. 
High scientific quality can be achieved even if the number of papers in the journal is not large and the impact relatively small. This happens for example in the cases of the \textit{Journal of Economic Literature} and \textit{Nature Immunology}.

\subsection*{Applicability to different impact metrics}

\begin{figure}
\centerline{\includegraphics[width=\figwidth]{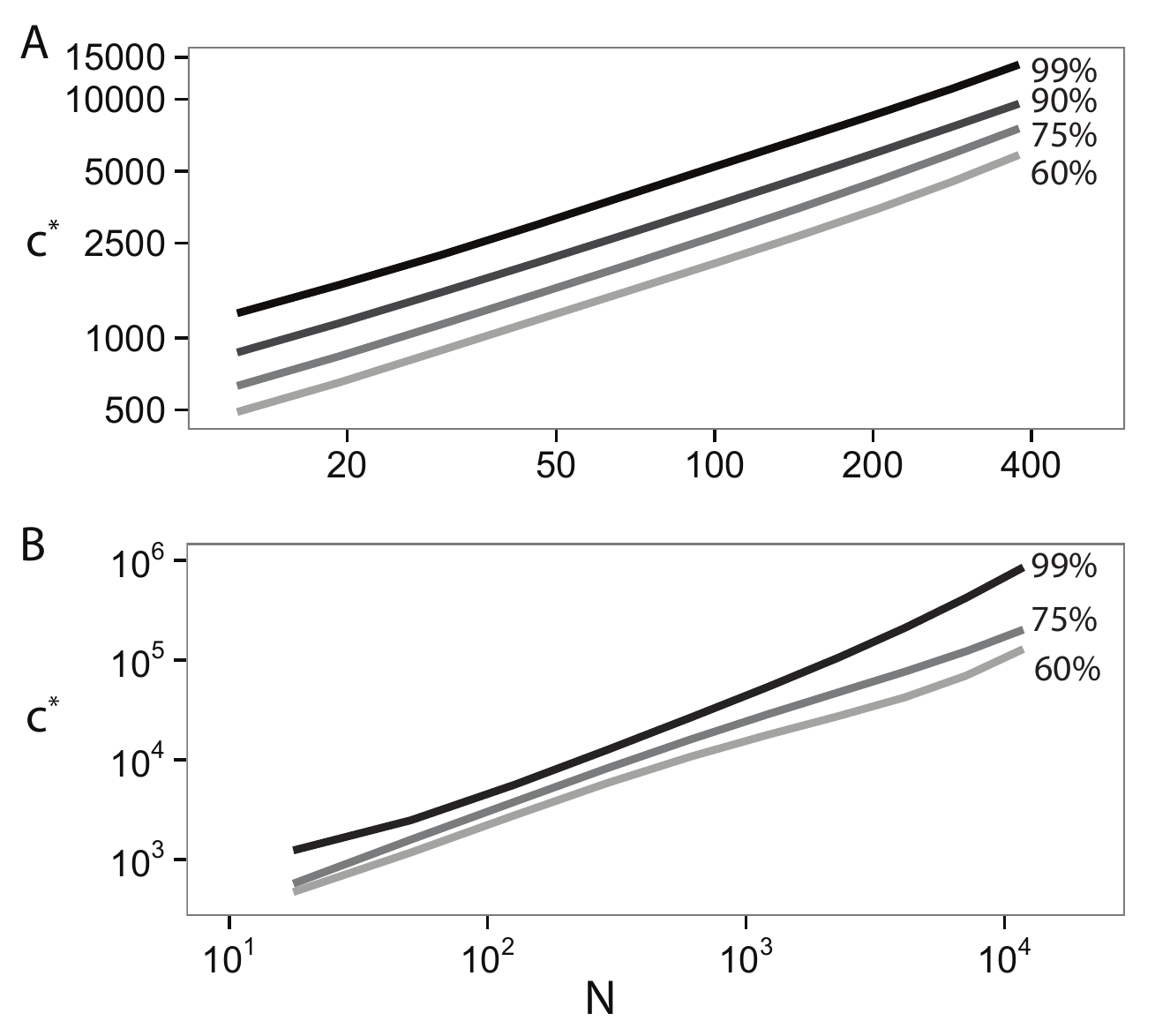}}
\caption{Relationship between critical $c^*$ and number of publications $N$ for (A) 996,288 authors and (B) 6,129 journals. Lines represent the values of $c$  above which a fraction $\delta$ of authors or journals in our dataset have $q>0.95$, for different values of $\delta$.}
\label{fig7}
\end{figure}

The proposed procedure can be used in conjunction with any arbitrary 
impact metric. As an illustration, let us consider the total number 
of citations $c$ in place of the $h$-index. 
Figure~\ref{fig7} plots the critical lines $c^*(N, \delta)$, defined analogously to $h^*(N, \delta)$, for authors and journals in the dataset. 
The $z$ scores measured on the basis of the two impact metrics, $h$ and $c$, are strongly correlated.

\section*{Discussion}

The role of citation-based metrics in the quantitative evaluation of research 
activities has become so central that numbers derived from bibliographic 
data influence the behavior of scholars and other stakeholders on a daily 
basis~\cite{king2004scientific, garfield2006history, kinney2007national}. 
Although the use of citation-based metrics as proxies 
for ``true'' scientific impact is still
debated~\cite{adler2009citation}, we believe that
many controversial issues associated with current evaluation practices can be
alleviated by designing better measure instruments.
Measurements are meaningful only 
if taken in reference to proper terms of comparisons.
Bibliometric numbers are instead often used as absolute quantities, 
and, as such, they do not convey much information.
While this issue can be easily addressed, at least when 
disciplinary biases are the concern, at the level of individual 
publications~\cite{radicchi2008universality, leydesdorff2011turning}, 
the proper evaluation of scholars  
represents a more challenging task~\cite{KRM}. Direct comparisons 
among individuals are not possible due to the intrinsic heterogeneity 
in publication records and career trajectories. Similar considerations are
valid for other aggregate entities, such
as journals, departments, and institutions, whose impact is generally quantified
with additive metrics over sets of publications. 

In this paper, we have radically changed the point of view 
of the methodology currently in use: the term of comparison 
of a researcher, journal or organization, is not
given by other real entities, but artificial copies of the same entity.
This procedure is very general, and allows us to assign a statistical 
significance to arbitrary impact metrics --- from simple ones, 
like the total number of citations, to more complex
ones, such as the $h$-index or the journal impact factor. Furthermore, 
the procedure can be applied to compensate for any arbitrary source of bias, 
in place of or in addition to disciplines and academic age. For instance, 
one could create clones by preserving types of publications, say 
journals versus conference proceedings, or countries, languages, and so on. 
The only statistical requirement is the availability of representative 
sets of publications in each category. 

We studied a large set of scientific
publications to show the utility of our approach in 
assessing scientific quality irrespective
of number of publications and impact. We demonstrated that the  
procedure is capable of singling out exceptional
journals and scholars. A natural extension of our study will be 
to perform the same quality analysis for even more heterogeneous entities, 
such as research groups and institutions.

An additional merit of the proposed evaluation system is to encourage 
parsimonious publishing strategies: 
increasing the number of publications also increases the 
critical threshold of impact necessary to demonstrate scientific quality. 
This will hopefully secure against the current way in which impact 
metrics are gamed by ``salami publishing,'' ``self-plagiarism,'' and 
``minimum publishable unit'' practices~\cite{Broad13031981}.

\section*{Acknowledgements}

The authors are grateful to the current and past members of the Center for Complex Networks and System Research at Indiana University (\url{cnets.indiana.edu}).
This work is supported in part by NSF (grants SMA-1446078 and CCF-1101743) and the J.S.~McDonnell Foundation (grant 220020274). The funders had no role in study design, data collection and analysis, decision to publish, or preparation of the manuscript. Authors declare no conflict of interests.


\bibliography{../quantity_quality}


\newpage


%
%
\begin{table}
\caption{Scientific quality of a sample of Nobel laureates. For each recipient we report the year of the award, the name of the laureate, the field of the award, and the $h$ and $z$ values associated with the academic profile.}
\begin{tabular}{cllrr}
\hline
Year & Laureate & Field & $h$-index & $z$-score\\
\hline
1991 & P.-G. de Gennes & Physics & 45 & 11.7\\
1991 & B Sakmann & Medicine & 103 & 20.0\\
1991 & E. Neher & Medicine & 87 & 15.2\\
1991 & R.H. Coase & Economics & 15 & 6.1\\
1998 & D.C. Tsui & Physics & 71 & 8.4\\
1998 & R.B. Laughlin & Physics & 51 & 5.1\\
1998 & H.L. Storrmer & Physics & 68 & 14.4\\
1998 & J.A. Pople & Chemistry & 109 & 23.7\\
1998 & W. Kohn & Chemistry & 42 & 5.7  \\
1998 & F. Murad & Medicine & 74 & 10.7\\
1998 & L.J. Ignarro & Medicine & 75 & 10.0\\
1998 & R. Furchgott & Medicine & 17 & 3.4\\
2012 & S. Haroche & Physics & 56 & 13.0\\
2012 & D.J. Wineland & Physics & 68 & 14.4\\
2012 & B.K. Kobilka & Chemistry &  73 & 13.8\\
2012 & J.B. Gurdon & Medicine & 56 & 6.2\\
2012 & L.S. Shapley & Economics & 10 & 2.1\\
\hline
\end{tabular}
\label{tab1}
\end{table}

\end{document}